\def\spose#1{\hbox to 0pt{#1\hss}}

\def\multleft#1{\hbox to size{\vbox {\halign {\lft{##}\cr #1}}\hfill}\par}
\def\multright#1{\hbox to size{\vbox {\halign {\rt{##}\cr #1}}\hfill}\par}

\def\today{\ifcase\month\or January\or February\or March\or April\or May\or
      June\or July\or August\or September\or October\or November\or December\fi
      \space\number\day, \number\year}
\def\s{\hbox{\phantom{5}}}	

\def\cm{{\rm\thinspace cm}}

\def\erg{{\rm\thinspace erg}}

\def\mJy{{\rm\thinspace mJy}}

\def\keV{{\rm\thinspace keV}}
\def\km{{\rm\thinspace km}}
\def\kpc{{\rm\thinspace kpc}}
\def\Lsun{\hbox{$\rm\thinspace L_{\odot}$}}

\def\Mpc{{\rm\thinspace Mpc}}
\def\Msun{\hbox{$\rm\thinspace M_{\odot}$}}

\def\s{{\rm\thinspace s}}
\def\ks{{\rm\thinspace ks}}
\def\yr{{\rm\thinspace yr}}

\def\ergps{\hbox{$\erg\s^{-1}\,$}}

\def\kmps{\hbox{$\km\s^{-1}\,$}}

\def\Msunpyr{\hbox{$\Msun\yr^{-1}\,$}}

\def\psqcm{\hbox{$\cm^{-2}\,$}}

\def\kmpspMpc{\hbox{$\kmps\Mpc^{-1}$}}

\def\H2{\hbox{H$_{2}$}}

\voffset- .65in

\documentclass[usegraphicx]{mn2e}

\usepackage{times}
\usepackage{amssymb}
\include{defn}

\begin{document}
\hsize=6truein

\title[XMM-Newton observations of two hyperluminous IRAS galaxies]{{\em XMM-Newton} observations of two hyperluminous {\em IRAS} galaxies: Compton-thick quasars with obscuring starbursts}
\author[R.J.~Wilman et al.]
{\parbox[]{6.in} {R.J.~Wilman$^{1}$, A.C.~Fabian$^{2}$, C.S.~Crawford$^{2}$ and R.M.~Cutri$^{3}$ \\ \\
\footnotesize
1. Sterrewacht Leiden, Postbus 9513, 2300 RA Leiden, The Netherlands. \\
2. Institute of Astronomy, Madingley Road, Cambridge, CB3 0HA. \\
3. IPAC, Caltech, MS 100-22, Pasadena, CA 91125, USA.\\ }}

\maketitle

\begin{abstract}
We present {\em XMM-Newton} observations of two hyperluminous {\em IRAS} galaxies ($L_{\rm{Bol}} > 10^{13}h_{\rm{50}}^{-2}$\Lsun), neither of which was previously detected by {\em ROSAT}. Published models of the infrared spectral energy distributions imply that a starburst and obscured quasar contribute equally to the power of each source.

IRAS F12514+1027 ($z=0.30$) is detected in 18.6~ks with 130 EPIC-pn counts over 0.2-12\keV. The soft X-ray spectrum exhibits thermal emission from the starburst, with $T \simeq 0.3$\keV~and $L(0.5-2.0\rm{\keV})=2.1 \times 10^{42}$\ergps. With its Fe K-edge, the flat continuum above 2\keV~is interpreted as cold reflection from a hidden AGN of intrinsic $L(2-10 \rm{\keV}) \gtrsim 1.8 \times 10^{44}$\ergps. Comparison with the infrared power requires that the X-ray reflector subtend $\sim 2\pi/5$~sr at the central engine. IRAS F00235+1024 ($z=0.575$) is not detected by the EPIC-pn in 15.9~ks; the limits imply that the starburst is X-ray weak, and (for the AGN) that any hard X-ray reflector subtend $<2\pi/5$~sr. The direct lines of sight to the AGN in both objects are Compton-thick ($N_{\rm{H}}>1.5 \times 10^{24}$\psqcm), and the presence of a reflection component in F12514+1027, but not in F00235+1024, suggests that the AGN in the latter object is more completely obscured. This is consistent with their Seyfert-2 and starburst optical spectra, respectively.
\end{abstract}


\section{INTRODUCTION}
About 50 galaxies are currently known with rest-frame 1--1000\micron~luminosities in excess of $10^{13}$\Lsun. These are the hyperluminous infrared galaxies (HyLIRGs), of which there are estimated to be 100--200 over the whole sky with $S_{\rm{60}}>200$\mJy~(Rowan-Robinson~2000; RR2000). The first were discovered during follow-up of the {\em IRAS} mission in the late 1980s, mainly through cross-correlation with existing optical, radio or active galaxy catalogues. Several have recently been identified from 850\micron~surveys, and the first sample based on unbiased infrared surveys is that of RR2000. 

Whilst the infrared emission in HyLIRGs is mainly thermally reprocessed dust emission, their ultimate power source -- and hence their relation to other galaxy populations -- has yet to be firmly established. One possibility is that they are simply high luminosity analogues of the ultraluminous infrared galaxies (ULIRGs, with $L_{\rm{IR}}>10^{12}$\Lsun), the majority of which are powered by obscured star formation, although above $L_{\rm{IR}}=4.5 \times 10^{12} h_{\rm{50}}^{-2}$\Lsun~the AGN-powered fraction is around 50 per cent (as inferred from mid-infrared {\em ISO} spectra: Lutz et al.~1998; Rigopoulou et al.~1999). Most ULIRGs reside in gas-rich interacting/merging galaxy systems, prompting a scenario in which they represent an early phase in the merger-induced formation of optically-selected quasars (see the review by Sanders \& Mirabel~1996). A different possibility is that the HyLIRGs constitute a population of primeval galaxies, undergoing their first major episode of star formation at rates $>10^{3}$\Msunpyr. This view was advocated by RR2000 on the basis of dust radiative transfer models of their spectral energy distributions (SEDs), from which he inferred that the bulk of HyLIRG emission longward of 50\micron~(in the rest frame) is starburst powered.

Further insights into the nature of the HyLIRGs have come recently from morphological analysis of their host galaxies and detailed SED modelling: Farrah et al.~(2002a) presented {\em HST} WFPC2 {\em I} band imaging of 9 HyLIRGs, three of which have interacting morphologies and six of which are QSOs. The QSO hosts are all extremely luminous ellipticals (with undisturbed morphologies) comparable to the brightest cluster ellipticals seen locally. None of the sources shows evidence for gravitational lensing, implying that their high luminosities are intrinsic. Verma et al.~(2002) presented {\em ISO} 7--180\micron~photometry and radiative transfer SED modelling of 4 HyLIRGs; they found that three of the objects require star formation (at rates $>3000$\Msunpyr) and AGN components to match their mid to far-infrared SEDs, whilst an AGN dust torus suffices for the fourth object. This modelling was extended by Farrah et al.~(2002b) with the addition of {\em SCUBA} sub-mm photometry for 11 HyLIRGs; they found that starburst and AGN components were required in all the objects, with the starburst fraction spanning the range 20--80 per cent (with a mean of 35 per cent). The AGN and starburst luminosities are correlated, but the trend of increasing AGN fraction with luminosity does not increase beyond that found for the brightest ULIRGs.

X-ray observations are an important diagnostic tool for HyLIRGs, particularly for revealing obscured AGN. Wilman et al.~(1998) presented {\em ROSAT} HRI observations of four HyLIRGs with narrow emission line optical spectra (two Seyfert 2s and two starbursts). None of the objects was detected, with a mean upper limit on $L(0.1-2.4\rm{\keV})/L_{\rm{Bol}}$ of $2.3 \times 10^{-4}$, implying that any AGN are either atypically X-ray weak or obscured by $N_{\rm{H}} > 10^{23}$\psqcm, or that these HyLIRGs are starburst dominated. IRAS F15307+3252 was also not detected to similarly stringent limits by {\em ASCA} (Ogasaka et al.~1997) and {\em ROSAT} (Fabian et al.~1996), despite evidence from spectropolarimetry for a hidden quasar nucleus (Hines et al.~1995). Conversely, the HyLIRG IRAS P09104+4109, situated in a cD galaxy at the centre of a rich cluster at $z=0.442$, has numerous X-ray detections, most recently by {\em Chandra} (Iwasawa, Fabian \& Ettori~2001): a strong 6.4\keV~iron K$\alpha$ line appears upon a reflected-dominated continuum from a Compton-thick obscured quasar; the directly-transmitted flux emerges above 30\keV~(in the rest-frame) and was detected by {\em BeppoSAX} (Franceschini et al.~2000). The recently discovered HyLIRG ELAIS J1640+41 at $z=1.099$ (Morel et al.~2001) is detected in the {\em ROSAT} all-sky survey with a flux consistent with its quasar optical spectrum.

In this paper we present {\em XMM-Newton} EPIC imaging spectroscopy of two HyLIRGs with {\em ROSAT} non-detections in Wilman et al.~(1998): IRAS F00235+1024 ($z=0.575$; $L_{\rm{Bol}}=1.85 \times 10^{13}h_{\rm{50}}^{-2}$\Lsun)  and F12514+1027 ($z=0.30$; $L_{\rm{Bol}}=1.51 \times 10^{13}h_{\rm{50}}^{-2}$\Lsun). The former has a starburst optical spectrum and {\em HST} imaging shows an interacting system (Farrah et al.~2002a). The latest SED models (Farrah et al.~2002b) imply that the starburst (with star formation rate $\sim 1000$\Msunpyr) and AGN contribute equally to the total infrared emission, and that the AGN dust-torus is seen almost edge on. IRAS F12514+1027 has a Seyfert 2 optical spectrum, but SED models suggest that the starburst and AGN again contribute almost equally to the total power. The bolometric luminosities for both objects are taken from Farrah et al.~(2002b) and are based on the latest SED modelling; they slightly exceed the values quoted by Wilman et al.~(1998). By probing to flux levels well below the {\em ROSAT} limits and at harder energies where photoelectric absorption is much lower, {\em XMM-Newton} offers the potential to constrain the properties of the putative active nuclei in these sources.

Throughout this paper we assume $H_{\rm{0}}=50$\kmpspMpc~and $q_{\rm{0}}=0.5$.

\section{OBSERVATIONS AND RESULTS}

\begin{figure}
\includegraphics[width=0.46\textwidth,angle=0]{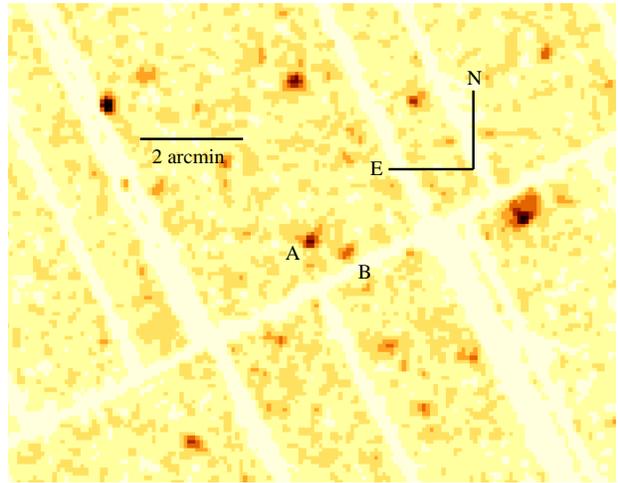}
\caption{\normalsize The {\em XMM-Newton} EPIC pn 0.2--12\keV~image of the field of IRAS F12514+1027. Source A (centroid at RA 12 54 0.90, Dec +10 11 13.3; J2000.0) is identified as the {\em IRAS} galaxy. No conclusion can be drawn on whether Source B is related to the HyLIRG. Each pixel measures 4.35 arcsec and the image has been smoothed with a gaussian of $\sigma=3.7$~arcsec.}
\label{fig:epicPNF12}
\end{figure}

\subsection{IRAS F12514+1027}
IRAS F12514+1027 was observed by {\em XMM-Newton} on 2001 December 28. The {\em EPIC}-pn detector was operated in Prime Full Window mode with the `thin' blocking filter, and the target placed near the aim point of CCD~1. The total pn observation time was 22.1\ks, but a background flare reduced the usable exposure to 18.6\ks~(of which 16.7\ks~is detector `live' time). The {\em IRAS} galaxy was not detected in the {\em EPIC} MOS cameras, so these data are not discussed further. The pn data were analysed with SAS version 5.2.0 on a Linux system. The {\em epchain} pipeline reduction was run and the {\em xmmselect} task used to produce an image in SKY coordinates in the 0.2--12\keV~band, part of which is shown in Fig.~\ref{fig:epicPNF12}. An X-ray source at RA 12 54 0.90, Dec +10 11 13.3 (J2000.0) is identified with the {\em IRAS} galaxy (source A in Fig.~\ref{fig:epicPNF12}). The HyLIRG is detected in the Two Micron All Sky Survey (2MASS) and is extremely red ($J=16.48 \pm 0.14$~mag, $H=15.21 \pm 0.11$~mag, $K_{\rm{S}}=13.52 \pm 0.05$~mag; $J-K_{\rm{S}}=2.96$). The {\em XMM-Newton} position of the HyLIRG is within 1.6\arcsec~of its 2MASS position of RA 12 54 00.81, Dec +10 11 12.3 (J2000). Since 2MASS positions are very accurate with respect to the Tycho 2 catalogue ($RMS \sim 0.1$\arcsec; H{\o}g et al.~2000) this illustrates the excellent positional reconstuction accuracy of {\em XMM-Newton}. The weaker X-ray source located $\simeq 40$~arcsec away at a position angle of $250$ degrees (source B) is not detected in the POSSII plates, and no conclusion can be drawn as to whether it is related to the HyLIRG.

For the X-ray analysis, events of PATTERN grades 1--4 (i.e. single and double pixel events) with quality FLAG 0 were selected to produce PI spectra from channels 0--20479, with a spectral binsize of 5 (for compatibility with the standard response matrices). A spectrum of IRAS F12514+1027 was extracted inside a circular region of radius 28~arcsec (limited in size by the presence of the `companion' object). According to the {\em XMM-Newton} Users' Handbook, the encircled energy fraction within this radius is around 80 per cent at 1.5\keV, rising to around 90 per cent at 10.5\keV. All luminosities quoted below have been corrected for source photons outside the extraction radius. A background spectrum was extracted from a source-free box measuring 1~arcmin on the side. There is known to be considerable spatial variation over the pn CCDs in the internal background from fluorescent lines of Al-K (1.4\keV) and the complex of Cu-K, Ni-K and Z-K lines at 7--9\keV~(see Lumb et al.~2002; Freyberg et al.~2002). Care was taken to ensure that the background region was sufficiently close to the source for this not to be a problem.

\begin{figure}
\includegraphics[width=0.46\textwidth,angle=0]{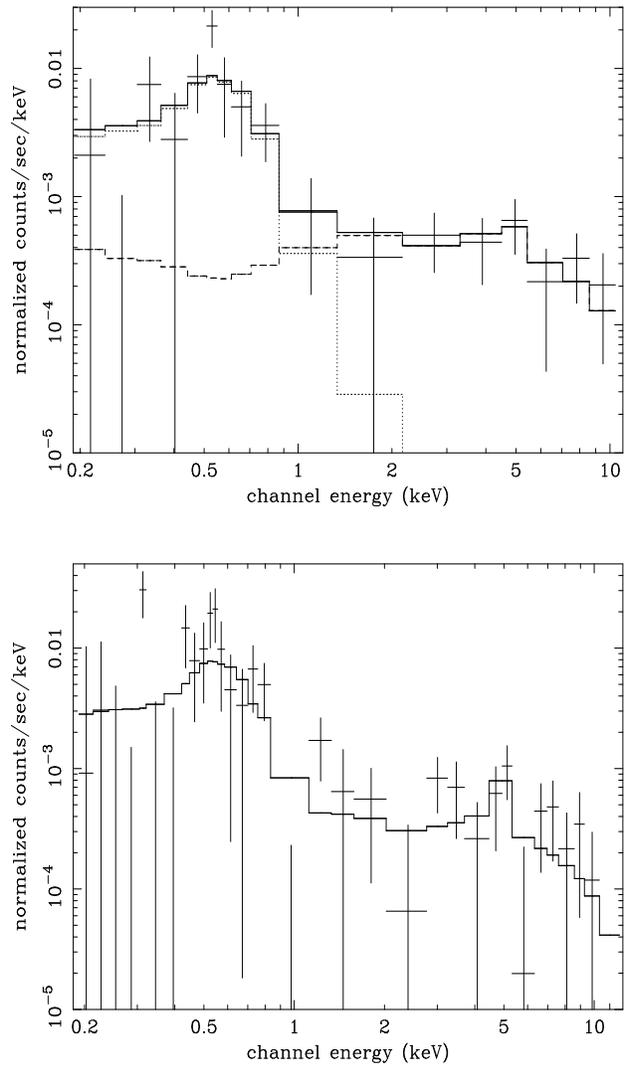}
\caption{\normalsize Upper: {\em EPIC}-pn spectrum of IRAS F12514+1027, grouped into a minimum of 20 counts per bin prior to removal of the background. The fit to the XSPEC model {\ttfamily zwabs(mekal) + pexrav}, described in the text, is shown by the solid line, (the individual model components are denoted by the dotted and dashed lines, respectively). Lower: As above but with a binning of 10 counts per bin and including a neutral 6.4\keV~Fe K$\alpha$ emission line in the model.}
\label{fig:F12514spec}
\end{figure}

The spectrum of IRAS F12514+1027 has $130 \pm 11$ net counts in the 0.2--12\keV~band and is shown in Fig.~\ref{fig:F12514spec}, grouped into a minimum of 20 counts per bin prior to background subtraction. An XSPEC model of the form {\ttfamily zwabs(mekal) + pexrav}~(convolved with the negligible Galactic absorption of $N_{\rm{H}}=1.73 \times 10^{20}$\psqcm; Stark et al.~1992) provides a good fit to the spectrum, with the parameters shown in Table~1. The thermal {\ttfamily mekal} component is probably of starburst origin, but we note, however, that the soft X-ray emission in starburst galaxies typically has a higher temperature of $\simeq 0.5-1.0$\keV~(see e.g. Iwasawa~1999; Dahlem, Weaver \& Heckman~1998). Together with the high intrinsic absorption required by this component -- which could imply an implausibly high mass of absorbing material if spatially extended on scales of tens of \kpc~-- this raises the possibility that the soft X-ray emission could instead be an emission line complex at $\simeq 0.8$\keV~from gas photoionised by the hidden AGN discussed below. However, the quality of the spectrum precludes further analysis of this possibility.

\begin{table}
\caption{Fit of the model {\ttfamily zwabs(mekal) + pexrav} to the spectrum of IRAS F12514+1027 (shown in Fig.~2, upper panel) }
\begin{tabular}{|ll|} \hline


Component & Parameters$^{\dagger}$ \\ \hline
{\ttfamily zwabs(mekal)} & $N_{\rm{H}}= 1.3^{+0.9}_{-0.7} \times 10^{21}$\psqcm \\
                         & $T=0.31^{+0.13}_{-0.05}$\keV \\ 
{\ttfamily pexrav} & For incident power-law: $\Gamma=2$ \\
                   & $E_{\rm{c}}=1000$\keV~(fixed) \\
		   & $refrefl=-1$~(i.e. reflection only) \\ 
		   & $cos i = 0.6$~(fixed) \\
{\ttfamily zwabs(mekal) + pexrav} & $\chi^2/\rm{d.o.f.} = 8.5/12$ \\ \hline

\end{tabular}
$\dagger$ Quoted errors are $1\sigma$
\end{table}

For the {\ttfamily pexrav} component (Magdziarz \& Zdziarski~1995), describing cold reflection from a hidden AGN, the best fit is obtained if only the reflection component is seen (i.e. no direct component is included). The photon index and exponential cut-off energy of the incident power law, and the cosine of the inclination angle of the reflector, are fixed to the values given in Table~1. There is evidence for a redshifted Fe K absorption edge at 5.5\keV~(7.1\keV~in the rest frame) but, surprisingly for a reflection-dominated source, no prominent Fe K$\alpha$ emission line at 6.4\keV~in the rest-frame (cf. the {\em Chandra} spectrum of IRAS P09104+4109; Iwasawa et al.~2001). When the spectrum is rebinned to a minimum of 10 counts per bin to enhance the sensitivity to any such line (Fig.~\ref{fig:F12514spec}, lower panel), a 90 per cent confidence upper limit on its intensity is found to be $1.1 \times 10^{-6}$~ph~s$^{-1}$~cm$^{-2}$, corresponding to a rest-frame equivalent width of 1.9\keV. This is in the expected range for a power-law spectrum incident on a medium with low ionization parameter in which the metals are neutral (as in the {\ttfamily pexrav} model) but no stronger inferences can be made (see e.g. Ballantyne, Fabian \& Ross~2002 and references therein).

For the assumed cosmology, the implied 0.5--2.0\keV~luminosity of the thermal component is $2.1 \times 10^{42}$\ergps, with no correction for intrinsic absorption (rising to $4.8 \times 10^{42}$\ergps~if absorption-corrected). For starburst galaxies $L(0.5-2.0)/L_{\rm{Bol}} \sim 10^{-4}$ (e.g. Kii et al.~1997; Iwasawa~1999), implying a bolometric luminosity for the starburst of $\sim 2.1 \times 10^{46}$\ergps~(assuming the observed luminosity), comparable with the figure of $7.90 \times 10^{12}$\Lsun~obtained by Farrah et al.~(2002b) from SED modelling. We stress, however, that the X-ray spectra of starbursts are more complex than just a galactic wind soft thermal component (Persic \& Rephaeli~2002); furthermore, this X-ray emitting gas accounts for only small fractions of the mass and energy of the wind, which can vary significantly from galaxy to galaxy (Strickland \& Stevens~2000). Caution is therefore required when using soft X-ray spectra to quantitatively classify starbursts. The {\em intrinsic} 2--10\keV~(rest-frame) luminosity of the AGN responsible for the reflection component is $1.8 \times 10^{44}$\ergps~if X-ray photons are reflected over a flat disc subtending 2$\pi$~sr; in such a highly obscured source the reflector is likely to have a somewhat more irregular geometry subtending less than 2$\pi$~sr, rendering this luminosity a lower limit on the true value. Since $L(2-10)/L_{\rm{Bol}}$ is around 0.03 for unobscured quasars (Elvis et al.~1994), the implied bolometric power of the AGN is $6 \times 10^{45}(2\pi/\Omega_{\rm{refl}})$\ergps~(where $\Omega_{\rm{refl}}$ is the solid angle subtended by the reflector) and certainly of quasar proportions.

The combined bolometric luminosity of the starburst and obscured quasar in this source is thus $[0.6 + 0.2(2\pi/\Omega_{\rm{refl}})]\times 10^{13}$\Lsun. For values of $\Omega_{\rm{refl}} \sim 2\pi/5$~sr, we can thus account for the entire bolometric luminosity of this HyLIRG. IRAS F12514+1027 therefore comprises a highly obscured (i.e. type II) quasar, together with a starburst of comparable luminosity. The starburst itself may constitute the obscuring medium, as proposed by Fabian et al.~(1998) to account for the high solid angle obscuration required by the hard X-ray background population.

\subsection{IRAS F00235+1024}
IRAS F00235+1024 was observed by {\em XMM-Newton} on 2001 January 10--11. The {\em EPIC}-pn detector was operated in Prime Full Window mode with the thin blocking filter, and the target placed near the aim point of CCD~4. A total pn exposure of 21.4\ks~was acquired, but a very strong background flare ($>10$ in count rate and probably due to soft protons -- see Lumb~2002) affected the first 5.5\ks~, resulting in a usable exposure of 15.9\ks~(of which 14.3\ks~is detector `live' time). 

\begin{figure}
\includegraphics[width=0.46\textwidth,angle=0]{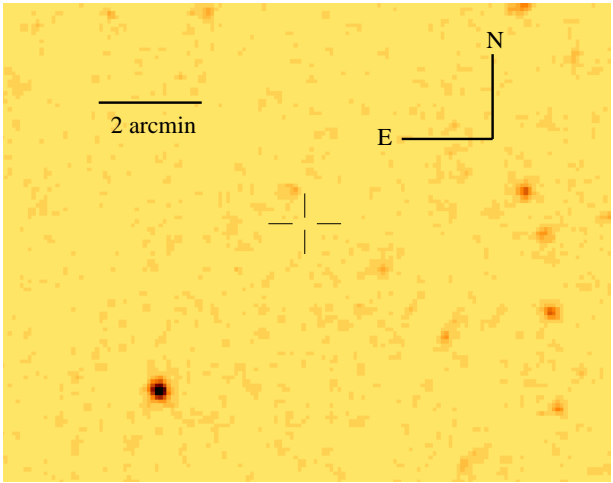}
\caption{\normalsize The {\em XMM-Newton} EPIC pn 0.2--12\keV~image of the field of IRAS F00235+1024. The cross-hairs mark the expected position of the IRAS galaxy (RA 00 26 06.7, Dec 10 41 27.6; J2000.0), showing that no X-ray source is detected here. Each pixel measures 4.35 arcsec and the image has been smoothed with a gaussian of $\sigma=3.0$~arcsec.}
\label{fig:epicF00235}
\end{figure}

The reduction followed exactly the same procedure as described above for IRAS F12514+1027. No source is detected at the position of IRAS F00235+1024 (RA 00 26 06.7, Dec + 10 41 27.6; J2000.0) in either the {\em EPIC}-pn or MOS cameras, but we confine our discussion to the more sensitive pn observations. Part of the pn image is shown in Fig.~\ref{fig:epicF00235}.

The upper limit on the number of counts in the 0.2--1.0\keV~band is 17 (determined as a $3 \sigma$ fluctuation of the background within a circular region of radius 35 arcsec). Assuming a thermal {\ttfamily mekal} model of $T=0.5$\keV~absorbed by the Galactic column density ($N_{\rm{H}}=5.1 \times 10^{20}$\psqcm; Stark et al.~1992), this translates into a $3\sigma$ upper limit on the rest-frame 0.5--2.0\keV~luminosity of such a component of $2.8 \times 10^{42}$\ergps, or $L(0.5-2.0)/L_{\rm{Bol}} < 4 \times 10^{-5}$. Given that SED models (Farrah et al.~2002b) imply that half the bolometric luminosity of this object originates in a starburst, this implies that for the starburst $L(0.5-2.0)/L_{\rm{Bol}}$ is slightly below the canonical value of $\sim 10^{-4}$ (Kii et al.~1997).

The $3\sigma$ upper limit on the number of source photons in the observed 2--10\keV~band is 19. Table~2 shows the corresponding upper limit on the intrinsic luminosity of any obscured quasar seen in transmitted light as a function of the absorbing column density (assuming a photon index of $\Gamma=2$ for the underlying power-law, typical of quasars). Since the SED modelling suggests the existence of an obscured quasar of $L_{\rm{Bol}}=9.1 \times 10^{12}$\Lsun, corresponding to $L(2-10\keV)=10^{45}$\ergps~(assuming as above the Elvis et al.~1994 hard X-ray--bolometric conversion), Table~2 implies that this quasar is completely obscured in direct light by a Compton-thick $(N_{\rm{H}} > 1.5 \times 10^{24}$\psqcm) absorber. Column densities of a few times this threshold are completely opaque to transmitted X-rays (see e.g. Wilman \& Fabian~1999). For a Galactic dust:gas ratio, the presence of a Compton-thick absorber in this source is consistent with the high inferred UV optical depth of the AGN torus ($\tau_{\rm{UV}}=1000-1500$; D.~Farrah, private communication). Assuming instead only reflected light (with the {\ttfamily pexrav} model as above), the implied upper limit on the {\em intrinsic} 2--10\keV~luminosity is $1.9 \times 10^{44}$\ergps~if the planar reflector subtends $2\pi$ sr at the illuminating source. Comparison with the intrinsic quasar luminosity implied by SED modelling implies that any scatterer must subtend $< 2\pi/5$~sr, or that any scattered X-rays from a more extended scatterer are subject to intrinsic absorption.

\begin{table}
\caption{Upper limit on the luminosity of any obscured quasar in IRAS F00235+1024 seen in direct light for various absorbing $N_{\rm{H}}$}
\begin{tabular}{|ll|} \hline
$N_{\rm{H}}$ (\psqcm) & L(2-10\keV)$^{\dagger}$ ($10^{43}$\ergps) \\ \hline
$10^{21}$ & $<1.6$ \\
$10^{22}$ & $<1.6$ \\
$5 \times 10^{22}$ & $<2.0$ \\
$10^{23}$ & $<2.4$ \\
$5 \times 10^{23}$ & $<6.3$ \\
$8 \times 10^{23}$ & $<10$ \\
$10^{24}$ & $<13$ \\ \hline
\end{tabular}

$\dagger$ Limits are $3\sigma$

\end{table}

\section{DISCUSSION}
IRAS F12514+1027 and F00235+1024 have comparable infrared dust luminosities  and SED modelling suggests that a starburst and obscured quasar contribute equally to the total power of each. Why then are their X-ray properties so different? 

In IRAS F12514+1027, we see the soft X-ray emission from the starburst at approximately the expected level. In the hard X-ray band ($E>2$\keV), the flat continuum and Fe K absorption edge suggest that the emission from the active nucleus is reflection dominated, implying that the direct line of sight to the nucleus is obscured by Compton-thick material. However, in order to account for the total luminosity, the X-ray reflector must subtend a solid angle $\sim 2\pi/5$~sr at the illuminating X-ray source. This relatively large opening angle for the escape of nuclear radiation is also consistent with the Seyfert 2 optical spectrum of this HyLIRG. 

In IRAS F00235+1024, the soft X-ray limit implies that the starburst is X-ray weak with respect to its infrared output (by at least a few tens of per cent compared with the `canonical' starburst). In the hard X-ray band, the direct line of sight to the active nucleus is obscured by Compton-thick material (as in F12514+1024 and consistent with the result from infrared SED modelling that the obscuring AGN torus is seen almost edge on; Farrah et al.~2002b). Unlike F12514+1024, however, no reflected X-ray emission is seen; any reflector must subtend a solid angle $<2\pi/5$~sr at the illuminating source. This is qualitatively consistent with the starburst optical spectrum of this object if the active nucleus is obscured over a large solid angle, so that line emission from stellar photoionization dominates over that from the active nucleus. If the merger-induced evolutionary scenario is applicable, this high level of nuclear obscuration may reflect the fact that F00235+1024 is a youthful merger, as shown in the {\em HST} image (Farrah et al.~2002a); no morphological information is available for IRAS F12514+1027 for comparison.

We conclude that in Compton-thick quasars such as these, it is the covering factor of the central engine which determines their appearance in reflected X-rays, and the AGN contribution to their optical spectra. Future hard X-ray observations of an enlarged sample of narrow optical emission line HyLIRGs can test the generality of this conclusion.

\section*{ACKNOWLEDGMENTS}
We thank the European Space Agency and associated organisations for the construction and successful operation of {\em XMM-Newton}. RJW acknowledges support from an EU Marie Curie Fellowship. ACF and CSC thank the Royal Society for financial support. Duncan Farrah, Kazushi Iwasawa and Takayuki Tamura are thanked for useful discussions. This publication makes use of data products from the Two Micron All Sky Survey, which is a joint project of the University of Massachusetts and the Infrared Processing and Analysis Center/California Insitute of Technology, funded by the National Aeronautics and Space Administration and the National Science Foundation.

{}

\end{document}